\documentclass{article}
\usepackage{amsmath}
\usepackage{array}
\usepackage{graphicx}
\usepackage{multirow}
\usepackage{adjustbox}
\usepackage[a4paper, total={6in, 8in}]{geometry}
\usepackage{authblk}
\usepackage{subcaption}
\usepackage{url}
\usepackage{rotating}
\usepackage{pdflscape}
\usepackage{color, colortbl}
\usepackage{array}
\usepackage{float}
\usepackage{stackengine}
\usepackage{enumitem}
\usepackage{hyperref}
\usepackage{orcidlink}
\setlist[itemize]{noitemsep}
\newcolumntype{?}{!{\vrule width 1pt}}

\newcolumntype{L}[1]{>{\raggedright\let\newline\\\arraybackslash\hspace{0pt}}m{#1}}
\newcolumntype{C}[1]{>{\centering\let\newline\\\arraybackslash\hspace{0pt}}m{#1}}
\newcolumntype{R}[1]{>{\raggedleft\let\newline\\\arraybackslash\hspace{0pt}}m{#1}}
\definecolor{Gray}{gray}{0.9}

\begin{document}
\title{Which topics are best represented by science maps? An analysis of clustering effectiveness for citation and text similarity networks}

\author{
Juan Pablo Bascur\orcidlink{0000-0002-4077-1024}
\and Suzan Verberne\orcidlink{0000-0002-9609-9505}
\and Nees Jan van Eck\orcidlink{0000-0001-8448-4521}
\and Ludo Waltman\orcidlink{0000-0001-8249-1752}
}

\date{June 2024}
\maketitle

\begin{abstract}
A science map of topics is a visualization that shows topics identified algorithmically based on the bibliographic metadata of scientific publications. In practice not all topics are well represented in a science map. We analyzed how effectively different topics are represented in science maps created by clustering biomedical publications. To achieve this, we investigated which topic categories, obtained from MeSH terms, are better represented in science maps based on citation or text similarity networks. To evaluate the clustering effectiveness of topics, we determined the extent to which documents belonging to the same topic are grouped together in the same cluster. We found that the best and worst represented topic categories are the same for citation and text similarity networks. The best represented topic categories are diseases, psychology, anatomy, organisms and the techniques and equipment used for diagnostics and therapy, while the worst represented topic categories are natural science fields, geographical entities, information sciences and health care and occupations. Furthermore, for the diseases and organisms topic categories and for science maps with smaller clusters, we found that topics tend to be better represented in citation similarity networks than in text similarity networks.
\end{abstract}

\section{Introduction} \label{Introduction}

Science maps \cite{chen2017science} are visualizations that provide an overview of the content of collections of scientific publications. The goal of science mapping is to find meaningful structures in the bibliographic metadata of publications (e.g, in the references, the titles and abstracts, or the authors). These structures can then be used for literature analysis or information retrieval \cite{cobo2011science,van2011methodological}. Some of the uses of science maps are field delimitation \cite{zitt2015meso}, research policy \cite{sullivan2007using}, and enhanced document browsing \cite{bascur2023academic}. A well established practice to create science maps is to cluster similar publications, and then to summarize the content of the resulting clusters. Our focus in this paper is on science maps created in this way.

When using science maps, it is important to be aware that scientific publications usually have more than a single topic (e.g., a document about \textit{lung cancer} is about both \textit{lungs} and \textit{cancer}), but in a science map they typically can be assigned to only one cluster, representing a single topic. Losing information when creating science maps is unavoidable, but it does raise the question of which of the topics addressed in a collection of publications a clustering will be based on. This is not an idle question, as there can be significant disagreement between expert-identified and cluster-identified topics \cite{held2021challenges}, indicating that expert-identified topics are poorly represented by the clusters in a science map. More specifically, an expert with an interest in a particular topic may find that publications related to this topic are scattered over many different clusters, with most of the publications in these clusters being unrelated to the expert's topic of interest. By providing a better understanding of the types of topics that are well or less well represented in science maps, we hope our research will contribute to a more effective use of these maps.

In this paper, we use the Medical Subject Headings (MeSH) terms to investigate clustering for biomedical topics. Our focus is on clustering solutions based on either citation or text similarity networks, which are the most common document similarity metrics for creating science maps. We aim to find out which MeSH terms are well represented by the clusters in a science map, a phenomenon that we will refer to as \textit{clustering effectiveness}. Our approach is to group topics, represented by MeSH terms, into topic categories, represented by branches of the MeSH tree, and to then evaluate clustering effectiveness at the level of these topic categories.

Our research questions are as follows:
\begin{itemize}
\item Which topic categories have the highest and lowest clustering effectiveness in citation and text similarity networks?
\item Which topic categories have higher clustering effectiveness in citation similarity networks than in text similarity networks, and vice versa?
\end{itemize}

In the remainder of this paper, we will discuss background literature, describe our data, define our metrics, report our analyses and discuss our results.

\section{Background}
This section has the following structure: In Subsection \ref{background:Evaluation of science maps} we explain how science maps are usually evaluated, in Subsection \ref{background:Criticism of science maps based on ground truth evaluations} we explore the criticism of science maps that originates from one particular evaluation method, and in Subsection \ref{background:Meaning of the clusters} we explain the challenges of understanding the meaning of the clusters in a science map.

\subsection{Evaluation of science maps} \label{background:Evaluation of science maps}
The most common method to evaluate the quality of a science map is to ask experts if the science map reflects their knowledge of the field of interest. The utility of this evaluation method has recently been called into question because it usually gives an inconclusive result: The experts tend to agree with most of the science map but identify caveats about certain details \cite{glaser2020opening}. Additionally, there are several issues intrinsic to the expert evaluation method: The evaluation criteria may differ between experts; seeing the map may affect the expert's understanding of a field; the expert may be biased towards the subfields of their interest; and the expert may have limited competence in some subfields \cite{glaser2020opening}.

An alternative method to evaluate the quality of a science map is to consider the intrinsic properties of the clustering process used to create science maps. Commonly used intrinsic properties are desirable characteristics such as homogeneous cluster sizes, few small clusters, stable clustering solutions between different runs of the cluster algorithm, and a short computing time to create the clusters \cite{vsubelj2016clustering}. An intrinsic properties evaluation method was developed by Waltman et al. \cite{waltman2020principled}. Their method assumes that there exists an ideal map and then assesses how closely a clustering solution matches this map. It evaluates the quality of a clustering solution based on one metric using another unrelated baseline metric (e.g., a clustering solution based on citation similarity can be evaluated using text similarity). Ahlgren et al. \cite{ahlgren2020enhancing}, who created the clustering solutions that we use in our current work, used this method with MeSH terms similarity as their baseline metric.

A third approach to evaluate the quality of a science map is to define a ground truth made of documents that correspond to a given topic, and evaluate the overlap between the clustering solution and the ground truth: either the extent to which all documents of each field are contained in a single cluster \cite{held2021challenges,held2020topic}, or the extent to which each cluster contains only documents of a single field \cite{rossetti2016novel,held2021challenges,held2022interpret,haunschild2018algorithmically}. Some studies obtained the ground truth from the references of review articles \cite{klavans2017type,sjogaarde2018granularity}, but most studies obtained the ground truth using expert knowledge. To our knowledge, MeSH terms have not been used as ground truths, although Sj{\"o}g{\aa}rde, Ahlgren and Waltman \cite{sjogaarde2021algorithmic} used MeSH terms to label clusters in science maps. It is worth mentioning that our work has a different goal than evaluating a science map based on a ground truth. Instead of evaluating the quality of a science map based on a set of topics, we evaluate which topics are most accurately represented in a science map.

\subsection{Criticism of science maps based on ground truth evaluations} \label{background:Criticism of science maps based on ground truth evaluations}
Evaluations that use expert knowledge ground truths have recently questioned the quality of science maps by challenging their ability to identify fields of science \cite{haunschild2018algorithmically,held2021challenges,held2020topic,held2022interpret}. For example, Held and Velden \cite{held2022interpret} found that science maps provide clusters about organisms rather than clusters about the field of invasive biology. One explanation for these negative results is that a document can belong to several fields or topics but only to a single cluster \cite{held2021challenges,held2022interpret} (although some maps allow documents to belong to multiple clusters \cite{xu2018overlapping,havemann2017memetic}). Another explanation is that the choice of a clustering algorithm can have a significant influence on the quality of a science map, and it is impossible to know beforehand which clustering algorithm will give the best result for a given map \cite{held2022know,rossetti2016novel}.

Similar negative findings have also emerged in areas beyond science mapping. For example, the field of complex systems has developed algorithms to clusters the elements that share a given property (i.e., the cluster matches the ground truth), but these algorithms fail in practical applications. On the other hand, this field has succeeded in practical applications of algorithms that infer the properties of an element based on the properties of the other elements in a cluster (e.g., fraud in telecommunications networks, function in biological networks) \cite{fortunato2010community,hric2014community, peel2017ground}.

\subsection{Meaning of the clusters} \label{background:Meaning of the clusters}
The negative findings discussed in the previous subsection suggest that science maps, and clustering in general, offer poor representations of certain ground truths. However, this does not mean that science maps are not useful. As mentioned in Subsection \ref{background:Evaluation of science maps}, experts tend to agree that science maps reflect their knowledge of a field. Also, in the field of complex systems, Newman and Clauset \cite{newman2016structure} argued that, even if clusters do not reflect the ground truth, they can still describe meaningful structures in the data. Our work tries to find out what kinds of structures are described by the clusters in a science map.

In this direction, Seitz et al. \cite{seitz2021case} found that the epistemic functions of citations (i.e., what kind of knowledge is a citation contributing to in a document) within a cluster are different from the epistemic functions of citations between clusters. This suggests that clusters tend to represent certain epistemic functions more than others. Also, the type of similarity network might have an effect on the meaning of clusters. For example, Ding \cite{ding2011community} found significant differences between clusters emerging from co-authorship networks of documents and clusters emerging from topic modeling of documents. On the other hand, Velden et al. \cite{velden2017comparison} found that there is a substantial similarity between the topics found in science maps built from citation and text similarity networks, although science maps built from citation networks are better at distinguishing topics when words related to the topics have multiple meanings.

\section{Methods}

This section has the following structure: In Subsection \ref{methods:dataselection}, we define how we selected our data. In Subsection \ref{methods:Data prepossessing}, we explain how we modified our data so to better fit our experimental design. In Subsection \ref{methods:Clustering effectiveness}, we explain how we evaluate the clustering effectiveness of topic categories.

\subsection{Data selection} \label{methods:dataselection}

\paragraph{Documents} The collection of documents that we use in our work comes from the work by Ahlgren et al. \cite{ahlgren2020enhancing}. This is a collection of 2,941,119 PubMed documents published between 2013 and 2017.

\paragraph{Clustering solutions} The clustering solutions that we use are the ones generated by Ahlgren et al. They created several clustering solutions for the above mentioned documents using different similarity metrics and granularities. They used the Leiden algorithm \cite{traag2019louvain} for clustering, where the parameter Resolution controls the granularity of the clustering solution (a higher Resolution value generates smaller clusters). We select two similarity metrics, one for citation and one for text, based on which pair of metrics produce similar cluster sizes at the same Resolution. The citation metric is \textit{Extended direct citation}, which is calculated using direct citations between documents plus the citations to documents outside the document collection \cite{waltman2020principled}. The text metric is \textit{BM25} \cite{INR-019}, which uses the noun phrases in the titles and abstracts of the documents, and weights them inversely to their frequency in the document collection \cite{waltman2020principled}. For each metric we selected the three clustering solutions that use the Resolution values $2*10^{-6}$, $2*10^{-5}$ or $2*10^{-4}$, enabling us to evaluate different cluster sizes. We selected these Resolution values because the first and second value yield cluster sizes similar to those in the algorithmic mapping of science \cite{waltman2012new} used in the CWTS Leiden Ranking \cite{Fields}, while the third value enables us to evaluate clusters of smaller size.

\paragraph{Topics} Our topics are the MeSH terms, a controlled vocabulary thesaurus from the National Library of Medicine (NLM) used for indexing PubMed. MeSH terms are semi-automatically annotated to documents by the NLM \cite{mesh}. We obtained the MeSH terms annotated for each document in our document collection, plus the metadata of the MeSH terms themselves, from the PubMed and MeSH databases (version from 2023) available in the database system of the Centre for Science and Technology Studies (CWTS) at Leiden University. 

\paragraph{Topic categories} Our topic categories are the 16 nodes at the first level of the MeSH hierarchical tree of topics \cite{mesh}, also known as the branches of the MeSH tree. We use branches because they group the MeSH terms in epistemological categories (e.g., organisms), which are the categories sometimes used for topical analysis of clusters \cite{held2021challenges,seitz2021case}. A single MeSH term can have instances in different branches of the MeSH tree. We will address this in Subsection \ref{methods:Data prepossessing}.

\subsection{Data prepossessing} \label{methods:Data prepossessing}

\paragraph{Clustering solution cleaning} We cleaned the clustering solutions by removing the clusters with fewer than 10 documents because these clusters usually had documents that were disconnected from the largest connected component of the similarity network. Removing these clusters removed only a minor fraction of the total number of documents. The statistics of each clustering solution after this process can be seen in Table \ref{table:Statistics of the clustering solution}. In this table, the variable $S$ is the smallest set of clusters that together cover at least half of the documents in the dataset. This means that $S$ contains the biggest clusters in the clustering solution. We report statistics for $S$ to provide some insight into the distribution of cluster sizes.

\begin{table}[t]
\caption{Statistics of the clustering solutions. $S$ is the smallest set of clusters that together cover at least half of the documents in the dataset.}
 \centering
 \includegraphics[width=0.5\columnwidth]{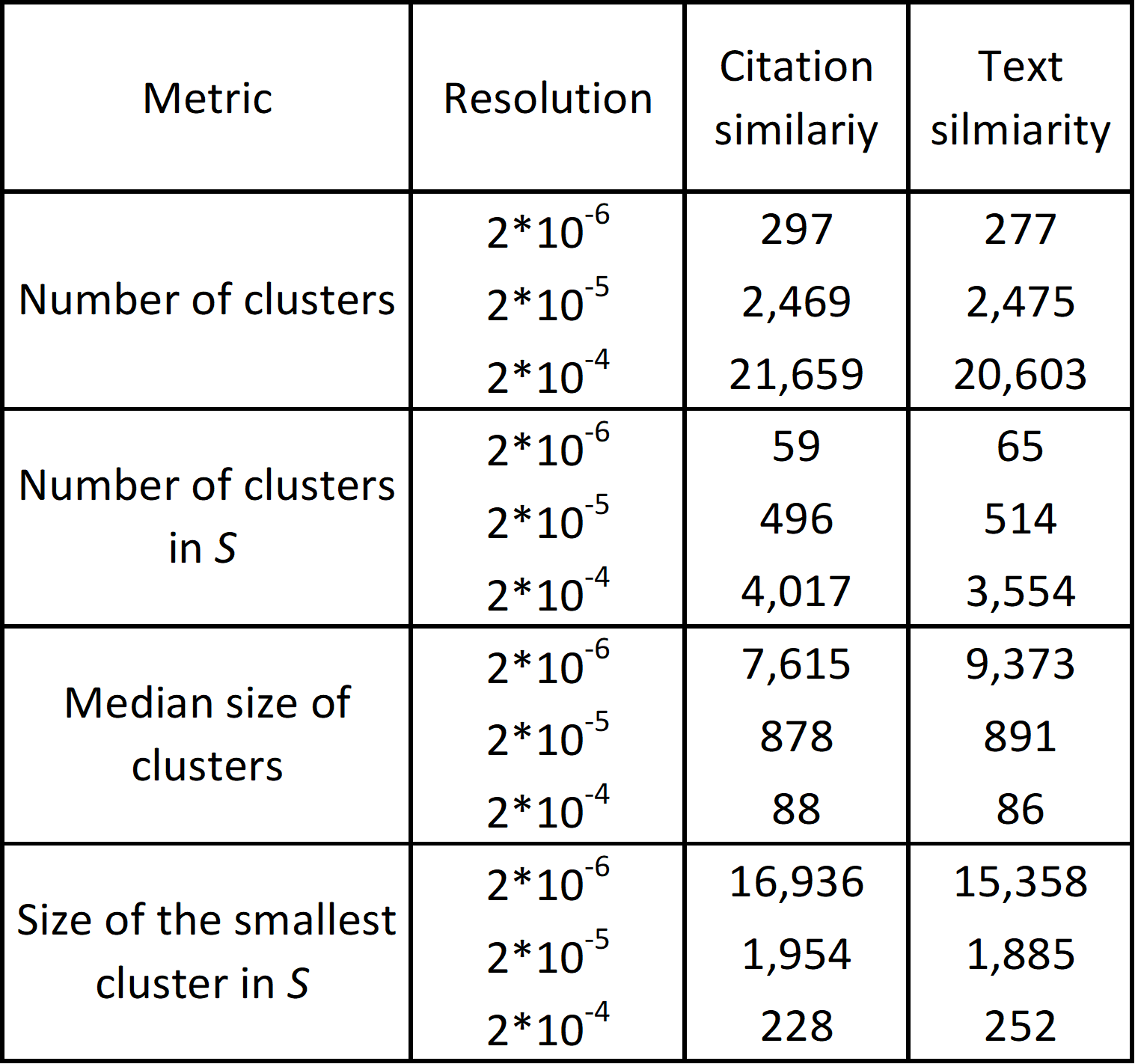}
\label{table:Statistics of the clustering solution}
\end{table}

\paragraph{MeSH term expansion} We would like a MeSH term to be annotated on all documents related to the topic of the MeSH term, but NLM typically only annotates up to 15 MeSH terms per document, which means that more generic MeSH terms are not annotated. To fix this, we expanded the number of MeSH terms annotated to a document by annotating, for each NLM MeSH term, all MeSH terms that are upstream in the MeSH tree, or in other words, all ancestors of the NLM MeSH term in the MeSH tree.

For example, if a document has the NLM MeSH term \textit{Abdominal Pain}, we also annotated the upstream MeSH term \textit{Pain}. While the former MeSH term belongs to the branch Diseases [C], the latter one belongs not only to the branch Diseases [C], but also to the branches Psychiatry and Psychology [F] and Phenomena and Processes [G]. We annotated the MeSH term \textit{Pain} paired with the branch Diseases [C], and not with the other two branches. For simplicity, in the rest of this paper we will refer to MeSH terms paired with a specific branch simply as MeSH terms. Also, we will refer to the documents that have a given MeSH term as the MeSH term documents and to the number of these documents as the MeSH term size.

\paragraph{MeSH term removal} We removed some MeSH terms to improve the quality of our experiments. Our first removal criterion is size. We removed MeSH terms with size greater than 300,000 (i.e., 10\% of the document set) because these MeSH term documents can saturate the clusters just by random chance, distorting our analysis. We also removed the MeSH terms with size 500 or less, because we want the smallest MeSH terms to be close but smaller than the median size of the clusters for resolution $2*10^{-5}$.

Our second removal criterion is redundancy. Due to the MeSH term expansion process, some MeSH terms had almost the same documents as their ancestor in the MeSH tree, like \textit{Dogs} and its ancestor \textit{Canidae}. This redundancy could distort our results. We therefore decided to remove the redundant MeSH terms by grouping together MeSH terms that share many documents and retaining only the smallest MeSH term from the group, which in our experience tends to be the term that best represents the group. The extent to which MeSH terms share documents was measured using Jaccard similarity, the grouping algorithm was agglomerate hierarchical clustering with the Complete Linkage method \cite{aggclus}, and the criterion for forming MeSH term groups was for MeSH terms to have a Jaccard similarity of at least 0.9. In cases where a group had more than one smallest MeSH term, we selected the one at the lowest level in the MeSH tree or the one with the largest number of instances in the MeSH tree.

\paragraph{Branch removal} To make our results more robust, we removed the branches with fewer than 100 MeSH terms. We ended up with the 14 branches shown in Table \ref{table:Number of MeSH terms}.

\paragraph{Size bins of MeSH terms} The size of a MeSH term can be expected to have an effect on its clustering effectiveness. We therefore grouped the MeSH terms according to their size. We refer to these groups as Size bins. To ensure the robustness of our results, we only considered Size bins that had at least 10 MeSH terms per branch. This resulted in five Size bins: 501-1,000, 1,001-2,000, 2,001-4,000, 4,001-8,000, and 8,001-16,000. The number of MeSH terms per Size bin can be seen in Table \ref{table:Number of MeSH terms}.

\begin{table}[t]
\centering
\caption{Number of MeSH terms per branch and Size bin.}
\includegraphics[width=0.90\columnwidth]{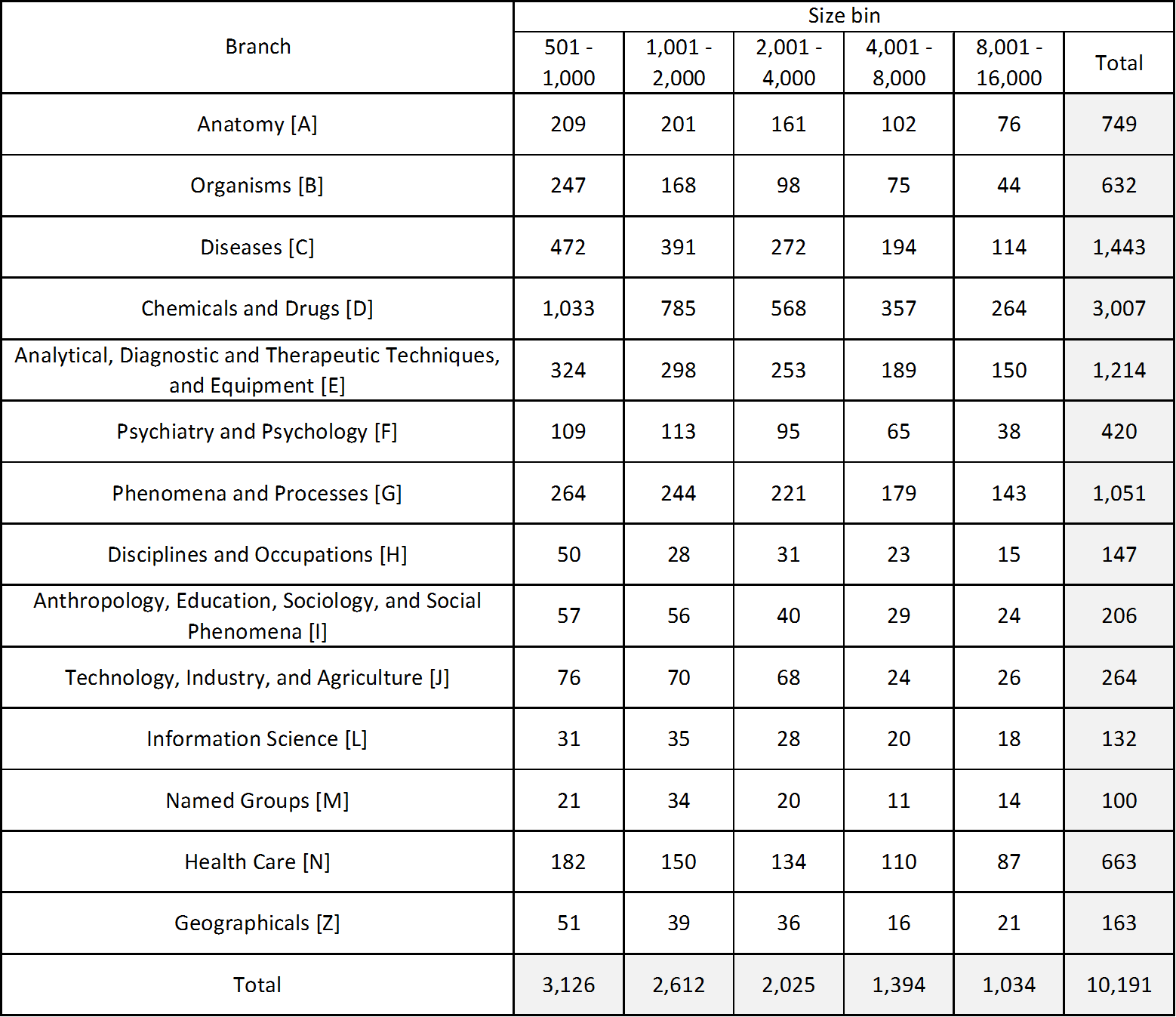}
\label{table:Number of MeSH terms}
\end{table}

\subsection{Clustering effectiveness} \label{methods:Clustering effectiveness}

\paragraph{Selection of clusters} To find out which MeSH terms are well represented by the clusters in a science map, we introduce the notion of clustering effectiveness. Measuring the clustering effectiveness of a MeSH term starts by selecting a subset of clusters. Our cluster selection criterion is to select the clusters with the largest number of MeSH term documents while making sure that the selected clusters cover at least a given share of all MeSH term documents. We call this share Coverage. We consider three Coverage values: 0.25, 0.50 and 0.75. Our cluster selection criterion minimizes the number of selected clusters for a given Coverage value. It is inspired by cluster quality metrics of Yuan, Zobel and Ling \cite{yuan2022measurement}. We expect our cluster selection criterion to reflect the clusters a user of a science map is likely to select while exploring the map.

\paragraph{Clustering effectiveness metrics} Once we have the selected clusters for a given MeSH term, we measure clustering effectiveness using two metrics:
\begin{itemize}
 \item Purity: Purity represents the extent to which the selected clusters are composed of MeSH term documents. It is the fraction of documents in the selected clusters that are MeSH term documents. In mathematical terms, Purity is defined as:
 \begin{equation}
 Purity = \frac{\sum_{i=1}^{N}{|D_i \cap D_{M}|}}{\sum_{i=1}^{N}|D_i|}
 \label{equation:purity}
 \end{equation}
Here, $N$ denotes the number of selected clusters, $D_i$ denotes the documents in selected cluster $i$ and $D_{M}$ denotes the MeSH term documents. The higher Purity, the more effective the clustering. Purity is bounded between zero and one.
 \paragraph{}
 \item Inverse count of clusters (ICC): ICC represents the extent to which the MeSH term documents are contained in a small number of clusters. ICC is defined as one divided by the number of selected clusters. In mathematical terms, ICC is defined as:
 \begin{equation}
 ICC = \frac{1}{N}
 \label{equation:icc}
 \end{equation}
The higher ICC, the more effective the clustering. Like Purity, ICC is bounded between zero and one.
\end{itemize}

We use two metrics instead of one to control for MeSH term size and cluster size: If there are few MeSH term documents, or if they are in big clusters, then ICC will be high but Purity will be low, and vice versa.

The Purity and ICC of a MeSH term are calculated for a given Coverage value, Resolution value and similarity network. We use C-Purity and C-ICC to refer to Purity and ICC calculated for a citation network, and T-Purity and T-ICC to refer to Purity and ICC calculated for a text network.

We also provide metrics for the difference in Purity and ICC between citation and text networks for a given MeSH term. These metrics, referred to as rPurity (Ratio Purity) and rICC (Ratio ICC), are calculated as the logarithm base 2 of C-Purity or C-ICC divided by T-Purity or T-ICC. In mathematical terms, rPurity and rICC are defined as:
\begin{equation}
rPurity = \log_2\left(\frac{\text{\textit{C-Purity}}}{\text{\textit{T-Purity}}}\right)
\label{equation:rpurity}
\end{equation}
\begin{equation}
rICC = \log_2\left(\frac{\text{\textit{C-ICC}}}{\text{\textit{T-ICC}}}\right)
\label{equation:ricc}
\end{equation}
Positive values indicate that a citation network yields a higher clustering effectiveness than a text network, and vice versa.

\section{Results} \label{Results}

\subsection{Which topic categories have the highest and lowest clustering effectiveness in citation and text similarity networks?}

To answer our first research question, we consider the C-Purity and T-Purity rankings of the 14 branches for each of the 45 combinations of parameter values (i.e., three Resolution values combined with three Coverage values combined with five Size bin values). Table \ref{table:ranking_purity} shows the number of times each branch appears in each position in the C-Purity and T-Purity rankings. The order of the branches in the table was determined manually so that the branches that frequently occupy higher ranking positions are above of the ones that occupy lower ranking positions. We found that the ICC rankings are strongly correlated with the Purity rankings, so we do not show them.

From Table \ref{table:ranking_purity} we make the following observations:
\begin{itemize}
 \item Most of the branches occupy between one and four adjacent positions, which shows that the position of the branches tends to be stable for different parameter values.
 \item For both C-Purity and T-Purity, the top five branches are almost always in positions 1 to 7, and the bottom four branches are almost always in positions 8 to 14. We therefore consider the top five and bottom four branches as the the ones with, respectively, the highest and lowest clustering effectiveness. 
 \item The top five and bottom four branches are the same for C-Purity and T-Purity, showing that in this respect citation and text networks yield very similar outcomes.
 \item The top five branches are Diseases [C], Organisms [B], Anatomy [A], Analytical, Diagnostic and Therapeutic Techniques, and Equipment [E] and Psychiatry and Psychology [F].
 \item The bottom four branches are Health Care [N], Disciplines and Occupations [H], Information Science [L] and Geographicals [Z].
\end{itemize}

\begin{figure}[t]
 \centering
 \includegraphics[width=0.9\columnwidth]{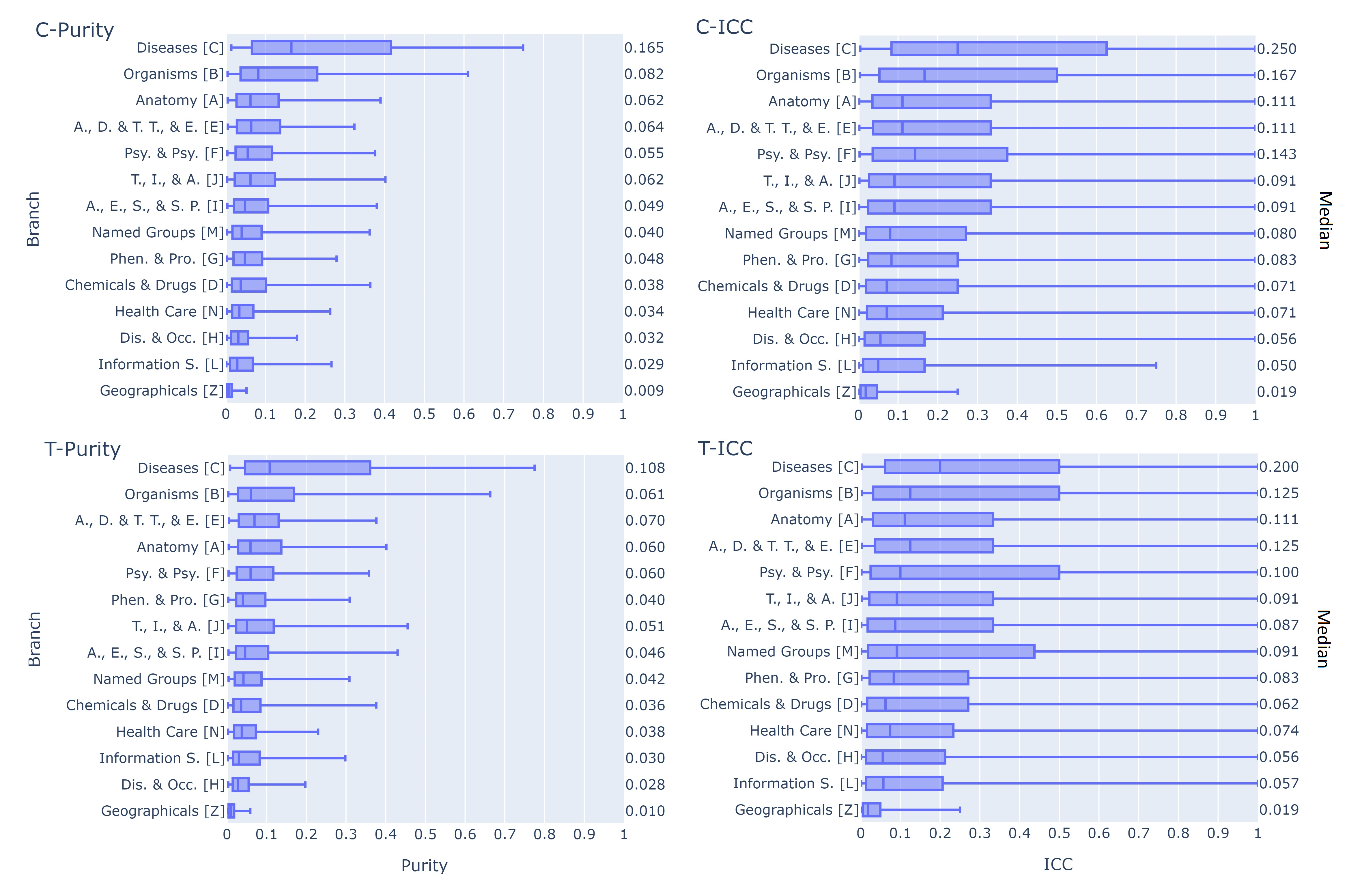}
 \caption{Box plots showing the distribution of C-Purity, C-ICC, T-Purity and T-ICC for each branch. The median values of each box plot are reported along the right Y axis. The branches are sorted as in Table \ref{table:ranking_purity}.}
 \label{figure:boxplot purity icc}
\end{figure}

Figure \ref{figure:boxplot purity icc} shows the distribution of the Purity and ICC values of each branch for the 45 combinations of parameter values. The box plots for the different branches heavily overlap with each other due to the effect of the parameter values on Purity and ICC. From Figure \ref{figure:boxplot purity icc} we observe that C-Purity, T-Purity, C-ICC and T-ICC are substantially higher for the branch Diseases [C] than for the other branches, while they are substantially lower for the branch Geographicals [Z]. This also explains why in Table \ref{table:ranking_purity} these branches almost always appear in position 1 and 14, respectively.

\begin{table}[t]
\caption{Number of times each branch appears in each ranking position, using either C-Purity (top) or T-Purity (bottom) as ranking criterion.}
\centering
\includegraphics[width=0.6\columnwidth]{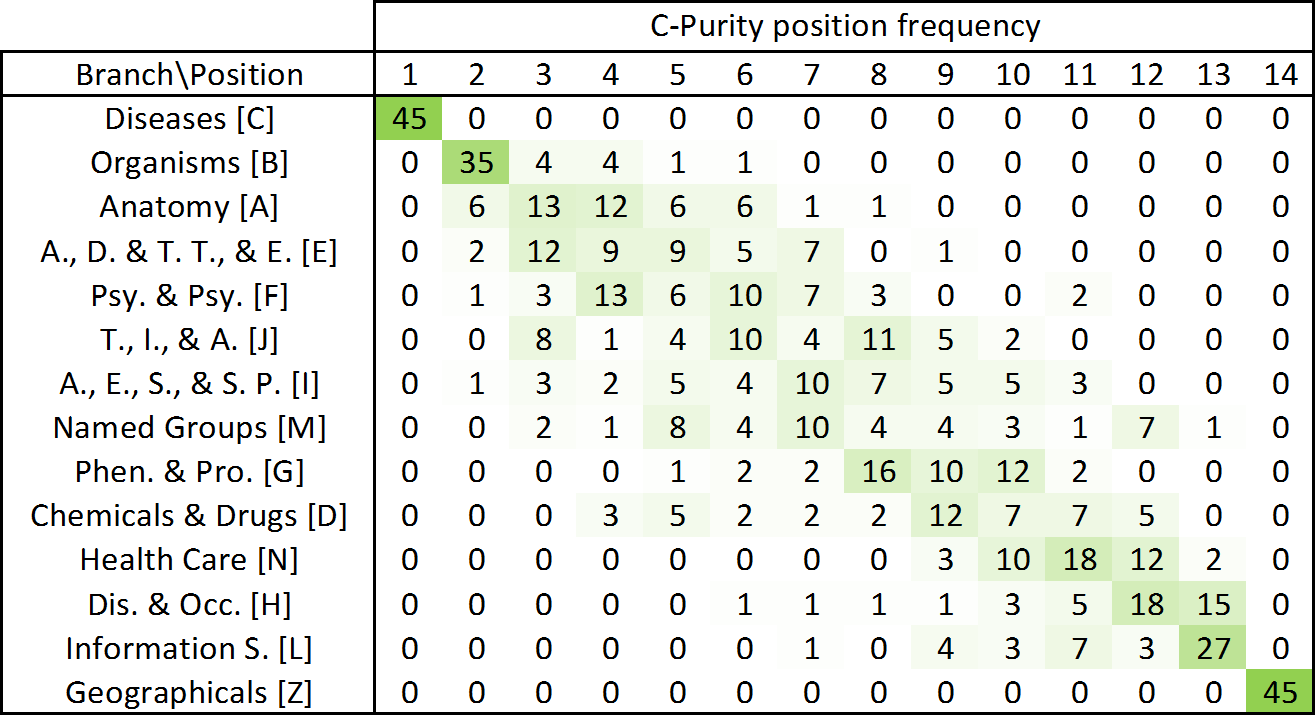}
\includegraphics[width=0.6\columnwidth]{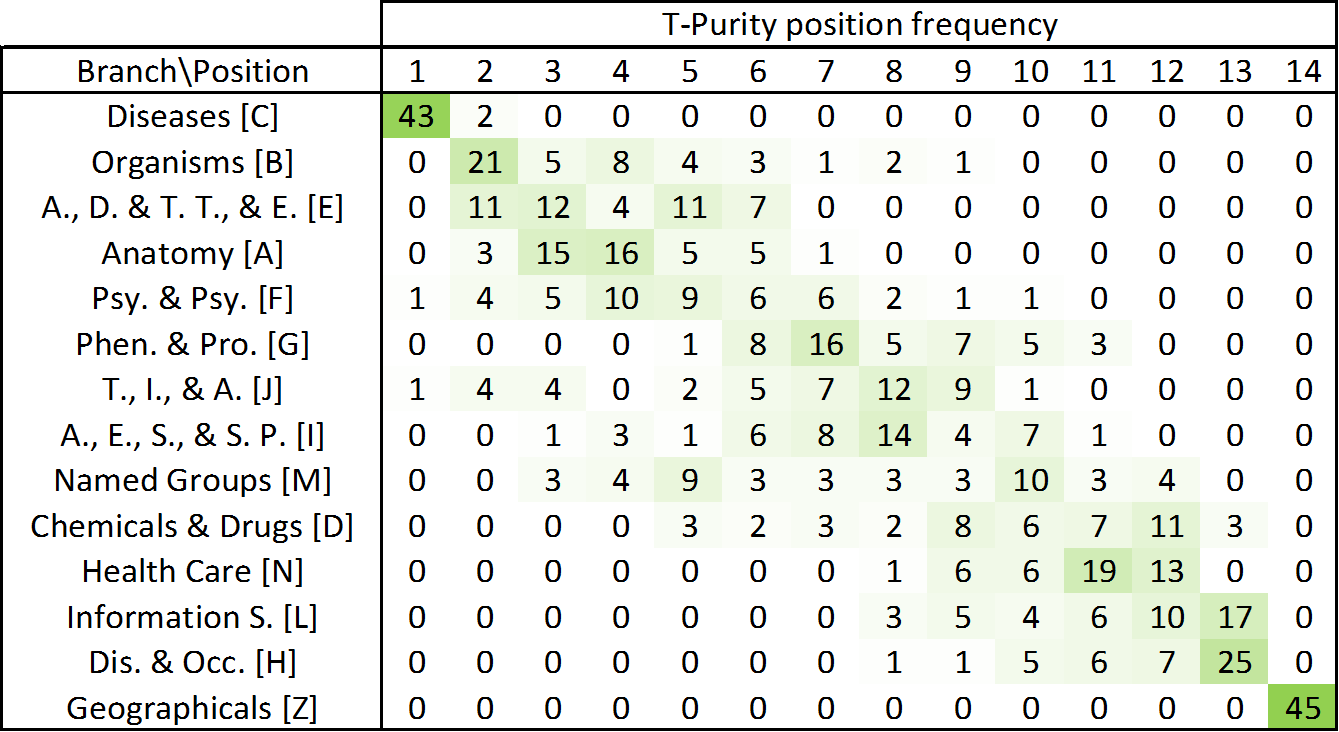}
\label{table:ranking_purity}
\end{table}

\subsection{Which topic categories have higher clustering effectiveness in citation similarity networks than in text similarity networks, and vice versa?}

\begin{figure}[t]
 \centering
 
 \begin{subfigure}[b]{0.49\columnwidth}
 \centering
 \includegraphics[width=1\columnwidth]{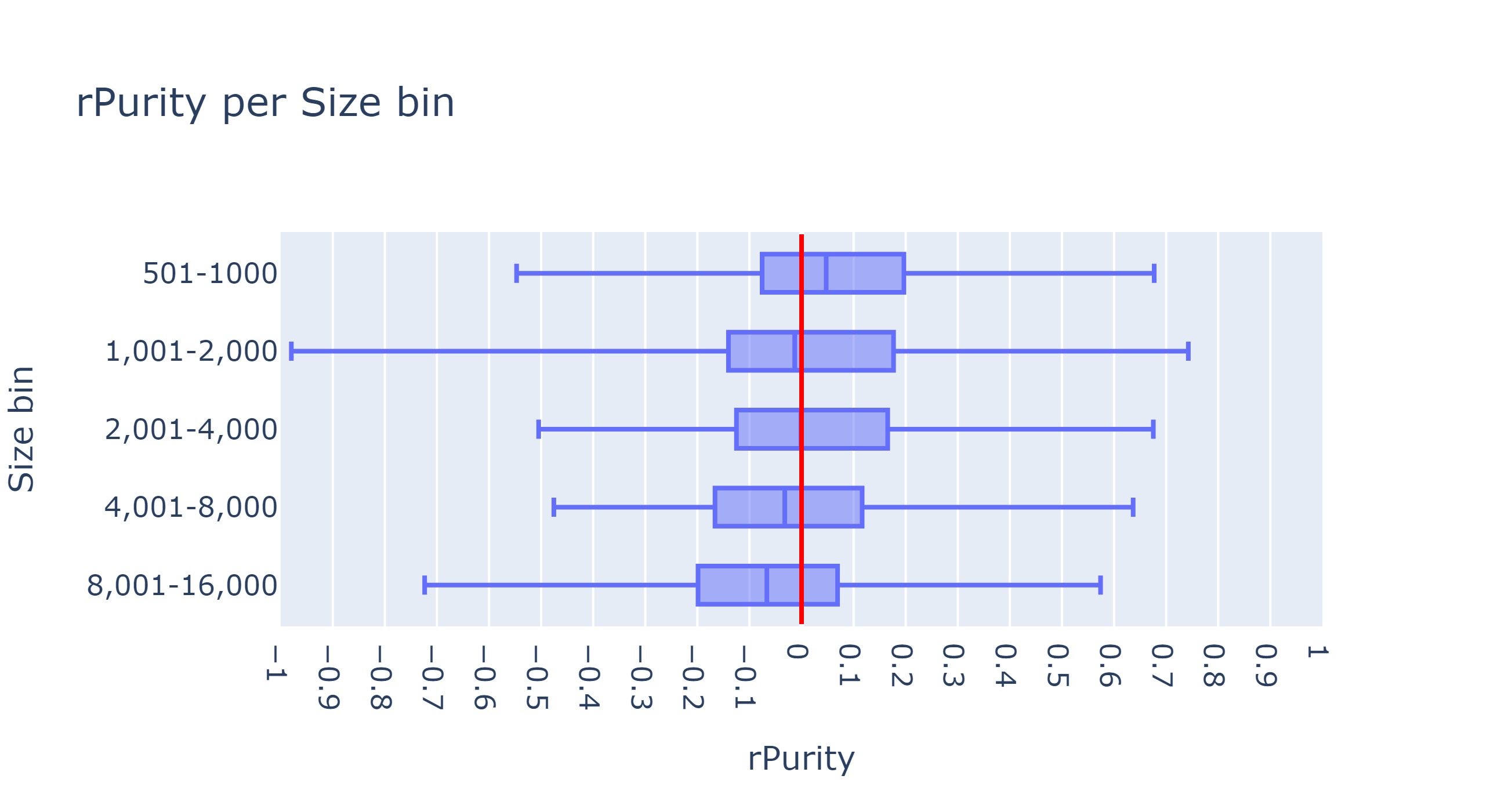}
 \end{subfigure}
 \hfill
 \begin{subfigure}[b]{0.49\columnwidth}
 \centering
 \includegraphics[width=1\columnwidth]{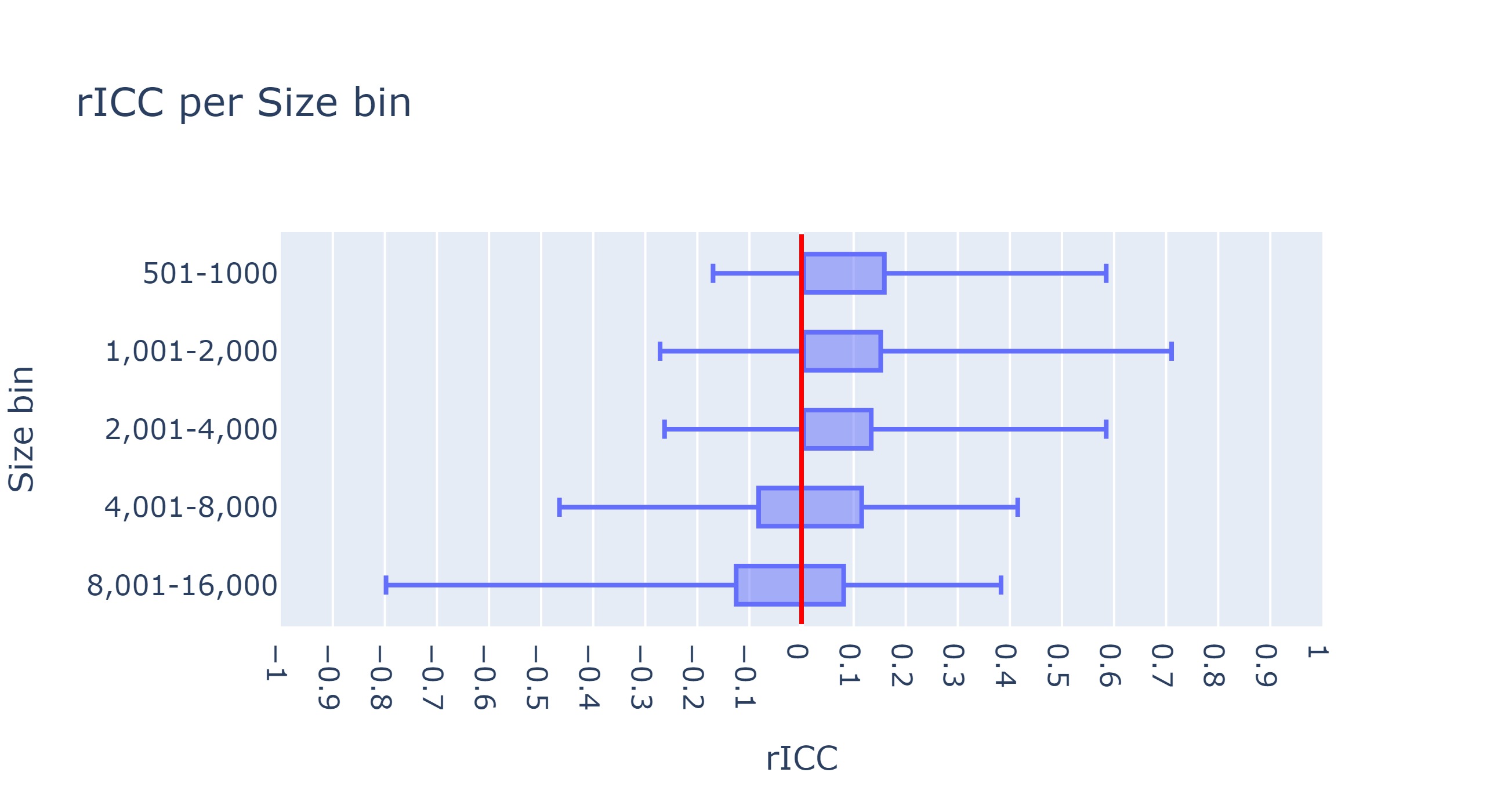}
 \end{subfigure}
 
 \begin{subfigure}[b]{0.49\columnwidth}
 \centering
 \includegraphics[width=1\columnwidth]{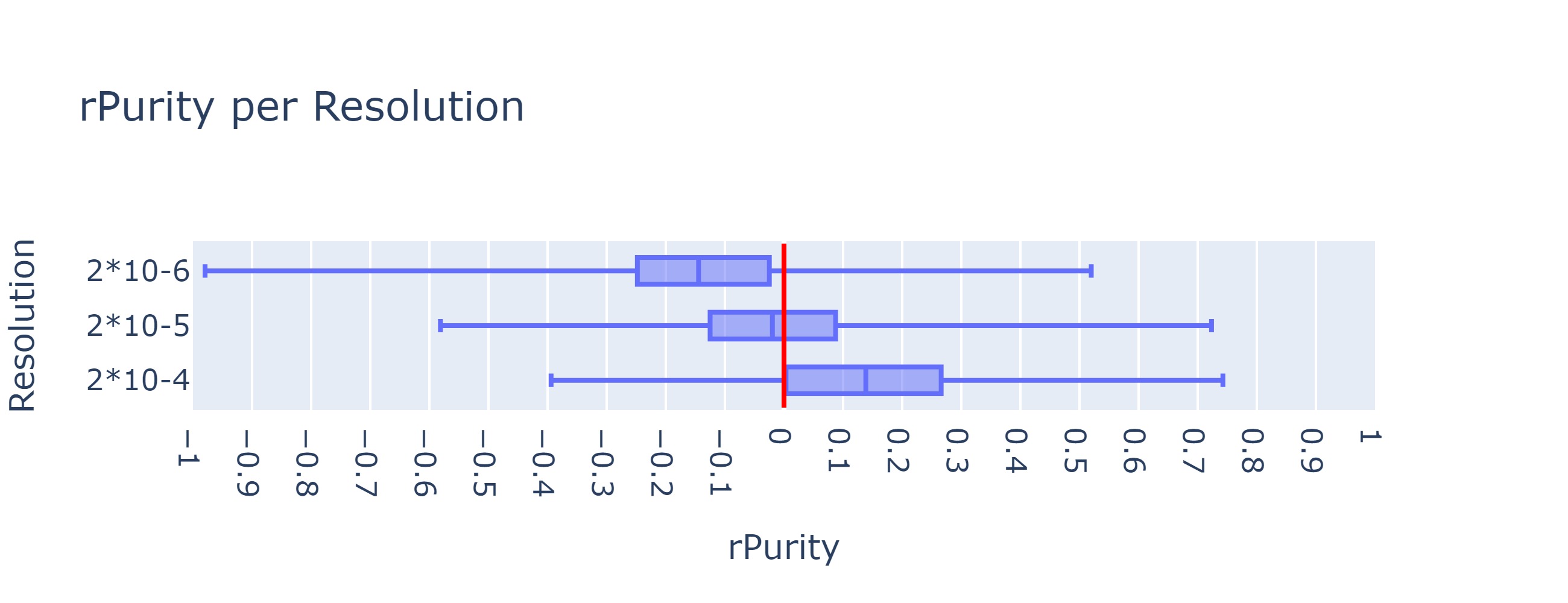}
 \end{subfigure}
 \hfill
 \begin{subfigure}[b]{0.49\columnwidth}
 \centering
 \includegraphics[width=1\columnwidth]{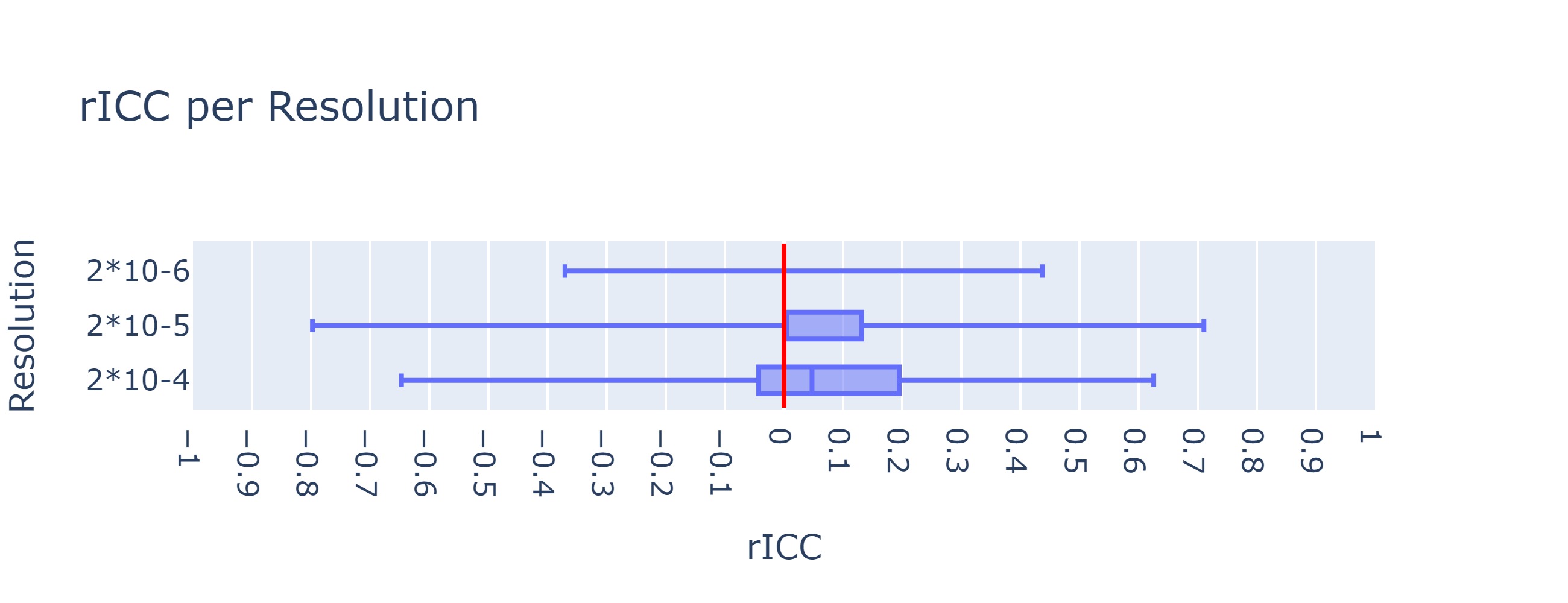}
 \end{subfigure}
 
 \begin{subfigure}[b]{0.49\columnwidth}
 \centering
 \includegraphics[width=1\columnwidth]{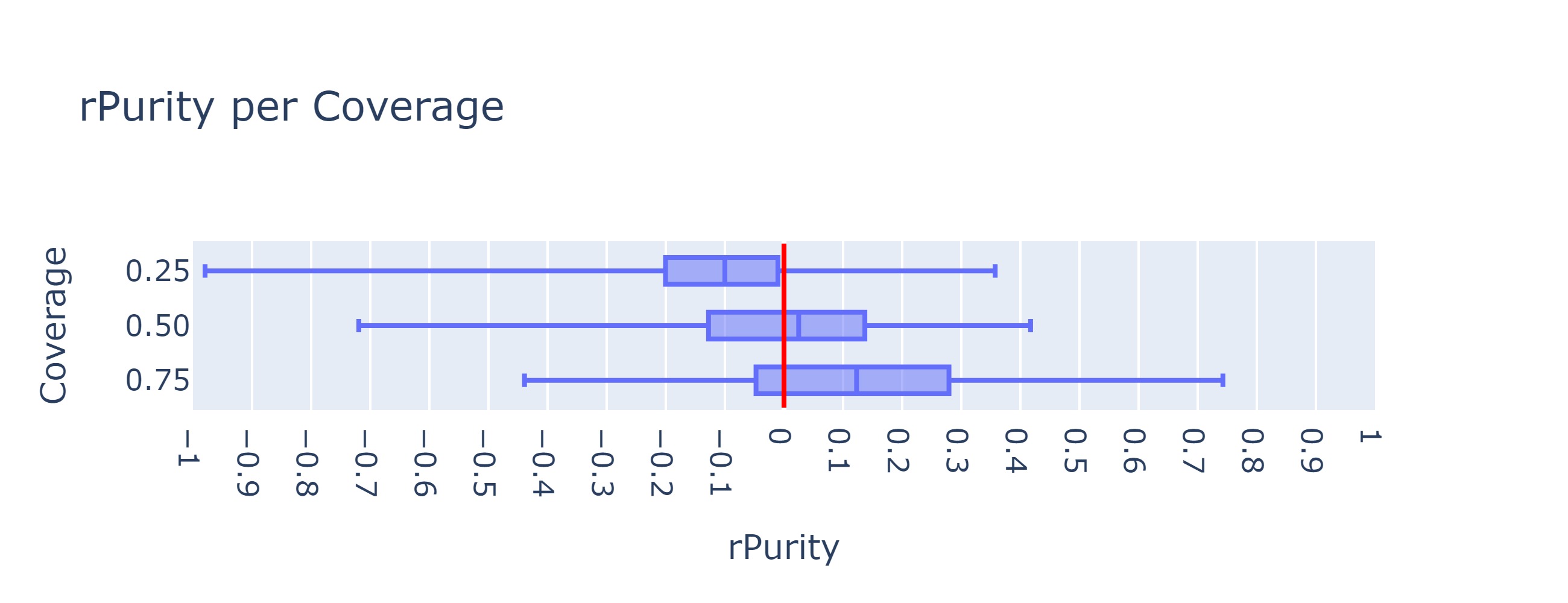}
 \end{subfigure}
 \hfill
 \begin{subfigure}[b]{0.49\columnwidth}
 \centering
 \includegraphics[width=1\columnwidth]{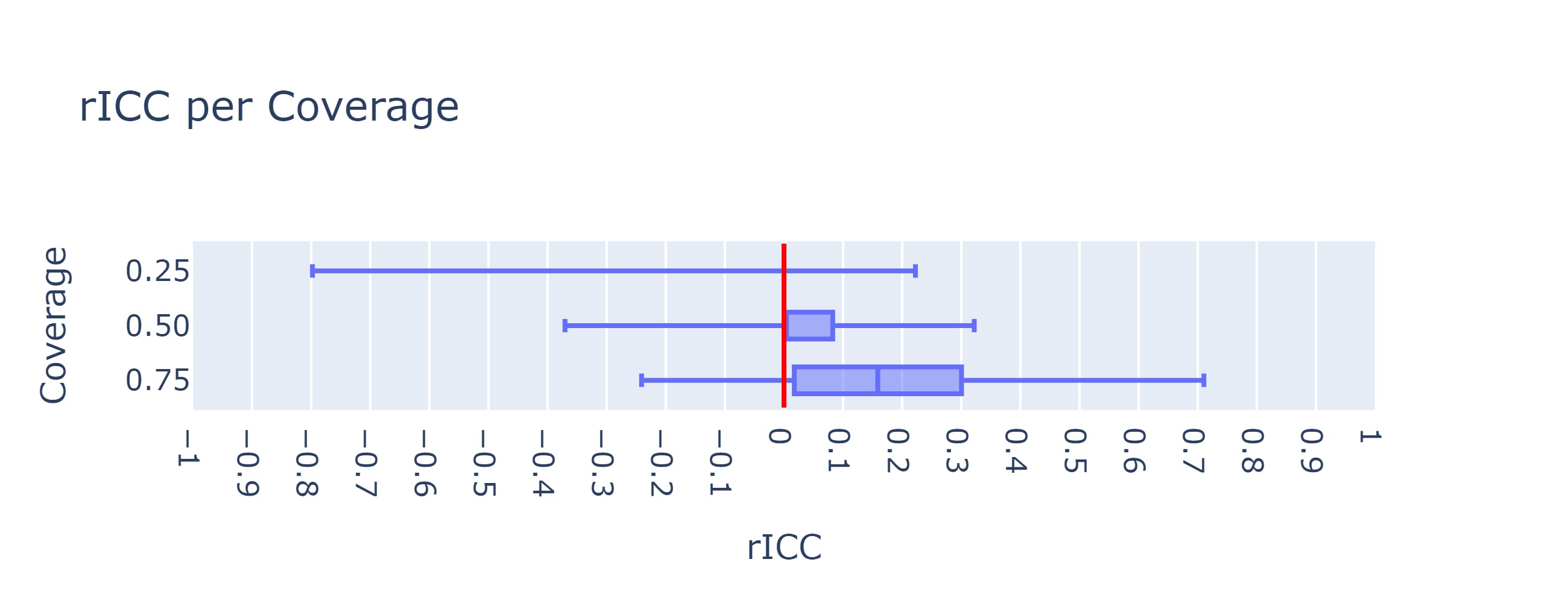}
 \end{subfigure}
 
 \caption{Box plots showing the distribution of rPurity and rICC for each value of Size bin, Resolution and Coverage.}
 \label{figure:ratio parameters}
\end{figure}

To address our second research question, we first evaluate how the ratio metrics rPurity and rICC correlate with the Size bin, Resolution and Coverage parameters. The box plots in Figure \ref{figure:ratio parameters} show the distribution of the rPurity and rICC values for each value of the Size bin, Resolution and Coverage parameters. Here we see that higher Resolution and Coverage are correlated with higher rPurity and rICC. Also, higher Size bin is correlated with lower rPurity and rICC, but this is a weak correlation.

\begin{figure}[t]
 \centering
 \begin{subfigure}[b]{0.49\columnwidth}
 \centering
 \includegraphics[width=1\columnwidth]{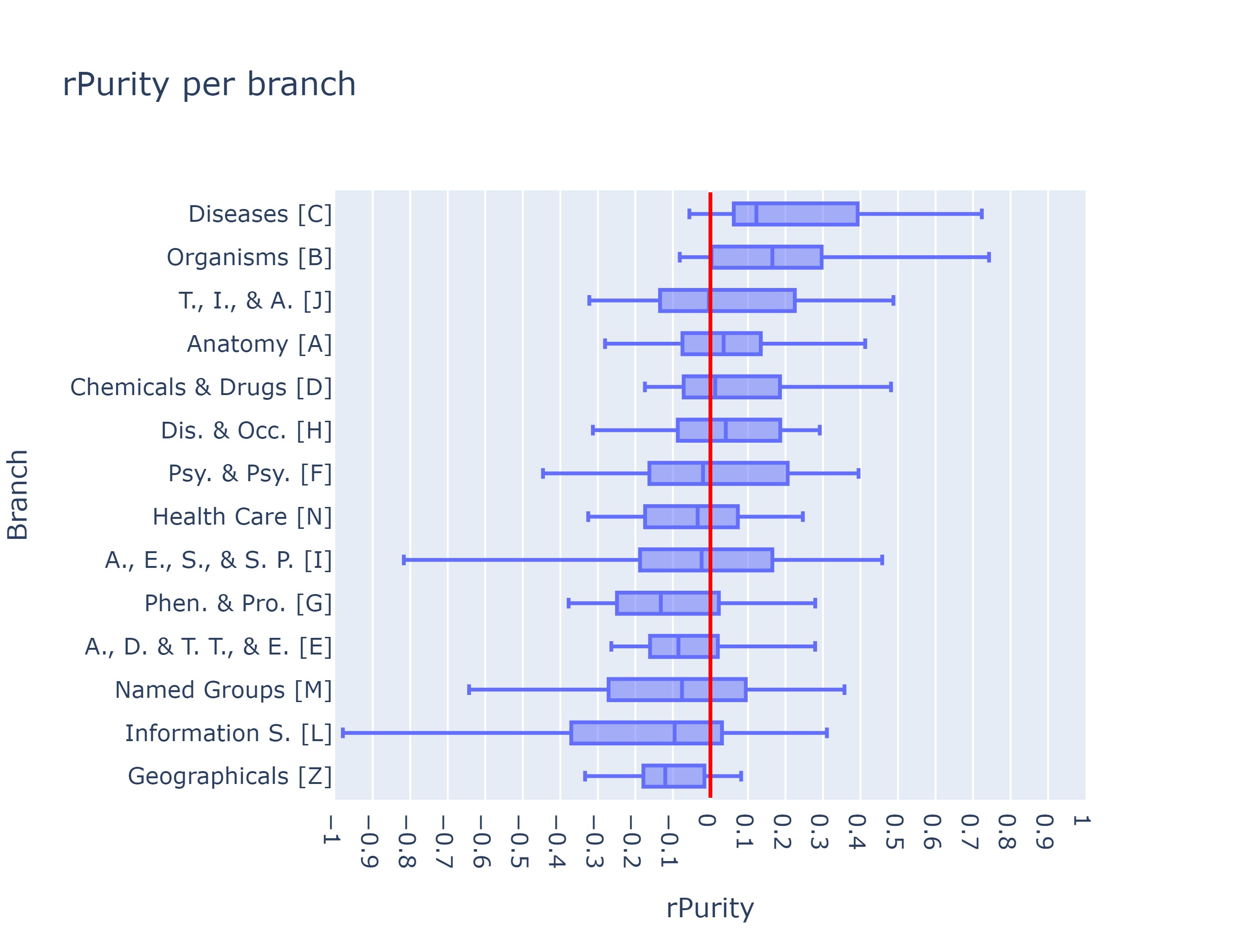}
 \end{subfigure}
 \hfill
 \begin{subfigure}[b]{0.49\columnwidth}
 \centering
 \includegraphics[width=1\columnwidth]{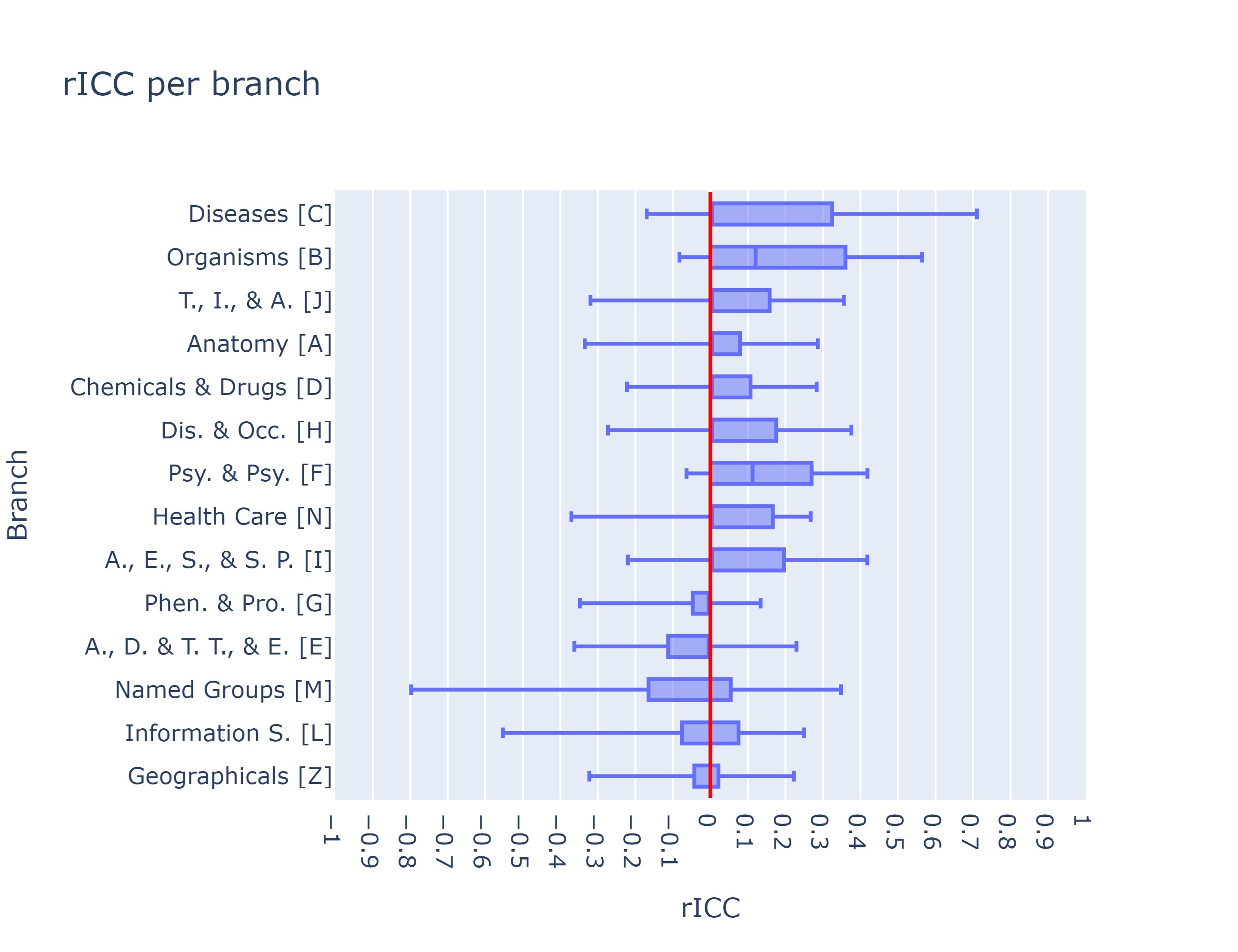}
 \end{subfigure}
 \caption{Box plots showing the distribution of rPurity and rICC for each branch.}
 \label{figure:ratio branches}
\end{figure}

The answer to our second research question depends on whether the rPurity and rICC values of a branch are positive or negative. Positive values indicate that the clustering effectiveness is higher in citation networks, while negative values indicate that the clustering effectiveness is higher in text networks. The box plots in Figure \ref{figure:ratio branches} show the distribution of the rPurity and rICC values of each branch for the 45 combinations of parameter values. For each branch, the rPurity and rICC distributions include both positive and negative values. This reflects the dependence of the rPurity and rICC values on the values of the Size bin, Resolution and Coverage parameters, as was shown in Figure \ref{figure:ratio parameters}.

Because for each branch the rPurity and rICC distributions include both positive and negative values, it is not possible to unequivocally conclude that a branch has a higher clustering effectiveness in either citation networks or text networks. Nevertheless, it is clear that the branches Diseases [C] and Organisms [B] tend to have a higher clustering effectiveness in citation networks than in text networks. rPurity and rICC are almost always positive for these branches. In contrast, the branches Geographicals [Z], Information Science [L], Named Groups [M], Analytical, Diagnostic and Therapeutic Techniques, and Equipment [E] and Phenomena and Processes [G] tend to have a higher clustering effectiveness in text networks than in citation networks. However, the results for these branches are less stable, so we need to be cautious in drawing strong conclusions.

\section{Discussion} \label{Discussion}

This section has the following structure: We discuss what we have learned for our first research question in Subsection \ref{Discussion:Which topic categories have the highest and lowest clustering effectiveness in citation and text similarity networks?}, for our second research question in Subsection \ref{Discussion:Which topic categories have higher clustering effectiveness in citation similarity networks than in text similarity networks, and vice versa?},  and for the strengths and weaknesses of our work in Subsection \ref{Discussion:Strengths and weaknesses}.

\subsection{Which topic categories have the highest and lowest clustering effectiveness in citation and text similarity networks?} \label{Discussion:Which topic categories have the highest and lowest clustering effectiveness in citation and text similarity networks?}

Our results show that the MeSH branches with the highest and lowest clustering effectiveness are the same for citation and text similarity networks. Despite the different purposes of writing and citing \cite{leydesdorff1998theories}, the way scientists write and the way they cite yield similar rankings of MeSH branches in terms of clustering effectiveness. It would be interesting to see if the top and bottom branches are also the same in other similarity networks, like co-tweeting \cite{costas2021heterogeneous}, co-authorship \cite{newman2004coauthorship}, and patent co-citation \cite{lai2005using}. 

The branch Disciplines and Occupations [H], which contains the MeSH terms for natural science fields, is among the branches with the lowest clustering effectiveness. This shows that how scientists cite each other is only weakly related to how they define scientific fields. This might suggest the need for alternative approaches to defining scientific fields, for instance based on science map clusters.

Held and Velden \cite{held2022interpret} reported that a given science map was poor at showing the field of invasive biology, and instead placed documents related to the field in clusters about species. Our results are in line with this, because invasive biology belongs to Disciplines and Occupations [H], one of the bottom four branches in our results, while species belongs to Organisms [B], one of the top five branches.

\subsection{Which topic categories have higher clustering effectiveness in citation similarity networks than in text similarity networks, and vice versa?} \label{Discussion:Which topic categories have higher clustering effectiveness in citation similarity networks than in text similarity networks, and vice versa?}

Our results show that which networks yield a higher clustering effectiveness depends strongly on the Resolution and Coverage values, with higher Resolution and higher Coverage increasing the clustering effectiveness for citation networks relative to text networks. Importantly, this does not mean that higher Resolution and higher Coverage increase the clustering effectiveness for citation networks in an absolute sense. It means that higher Resolution and higher Coverage increase the ratio between the clustering effectiveness for citation networks and the clustering effectiveness for text networks.

Ahlgren et al. \cite{ahlgren2020enhancing} developed a method to measure the accuracy of the clusters in a science map. Using their data and visualization method, we found that the accuracy of citation networks relative to text networks increases as the Resolution value increases. This is in line with our results. Unfortunately, we do not know the mechanism behind this dependency. Our findings for Resolution could be useful for users of science maps: It tells them that, if they have two science maps, one based on citations and another based on text, then decreasing the size of the clusters will make the citation one more effective relative to the text one, and vice versa.

In the context of field delimitation tasks, where a user of a science map identifies the clusters that contain the documents of a field, Coverage is analog to the completeness of the field delimitation. Our findings for Coverage suggest that citation networks are better for exhaustive field delimitation, while text networks are better for less exhaustive field delimitation.

Our results also indicate that, omitting the effect of Resolution and Coverage, the branches Diseases [C] and Organisms [B] tend to have higher clustering effectiveness in citation networks than in text networks. To exemplify what this means for users, we consider the use case of Held and Velden \cite{held2022interpret} discussed above: They would like to have a clustering of the field of invasive biology, but in their science map invasive biology documents are spread over clusters about organisms. If instead of a citation network a text network is used, the organisms will probably be clustered less effectively, which may give the opportunity for invasive biology documents to form their own clusters instead of being part of clusters about organisms.

\subsection{Strengths and weaknesses} \label{Discussion:Strengths and weaknesses}

We see the use of MeSH terms as an important strength of our work. An alternative approach could be to ask experts to assign documents to topics, but this cannot be done at the scale at which MeSH terms provide document-topic links. Also, MeSH terms link documents to topics at a scale that no other classification scheme, like the Mathematics Subject Classification, the ACM Computing Classification System, or the Physics Subject Headings, is able to provide.

We also improved the utility of the MeSH terms by using Coverage, MeSH term expansion, MeSH term removal and MeSH branches in our experimental design. Coverage diminished the effect of mislabeled documents (e.g., the document with DOI \textit{10.1007/s12603-020-1457-6} is incorrectly labeled with the MeSH term \textit{Alcohol Drinking}) by ignoring a certain share of the documents with a particular MeSH term. MeSH term expansion allowed us to have a collection of documents for each MeSH term that represent the topic of the MeSH term more accurately. MeSH term removal allowed us to ensure that our results are not affected by redundant MeSH terms. Using the MeSH branches as topic categories allowed us to use a curated scheme of topic categories. However, some topic categories may be absent from the MeSH tree (e.g., topics linking diseases with their medicines) and some lower levels of the MeSH tree may be more informative as topic categories (e.g., the children of the branch Disciplines and Occupations [H] are \textit{Natural Science Disciplines} and \textit{Health Occupations}, which may be more informative as topic categories than the branch itself).

Another strength of our work is that we evaluated clustering effectiveness per MeSH term, while other studies, like Waltman et al. \cite{waltman2020principled}, evaluated a clustering solution as a whole. Our method is also insensitive to the effect of size differences between MeSH terms and clusters (e.g., if clusters are much bigger than MeSH terms, it is impossible to have maximum Purity, and if they are much smaller, it is impossible to have maximum ICC) because our focus is on comparing the clustering effectiveness of different topic categories instead of achieving optimal clustering effectiveness.

A weakness of our work is that we used only one clustering algorithm, the Leiden algorithm, an algorithm that is commonly used by the science mapping community. Other studies used multiple algorithms: Held, Laudel and Gläser \cite{held2021challenges, held2020topic} analyzed clusters created by the Leiden algorithm and the Infomap algorithm. Held \cite{held2022know} assessed the suitability of the Leiden, Louvian, OSLM and Infomap algorithms for creating clusters. Beyond science maps, Rossetti, Pappalardo and Rinzivillo \cite{rossetti2016novel} showed that different clustering algorithms (Louvain, Infohiermap, cFinder, Demon, iLCD and Ego-Network) have differential performance for different types of networks (DBLP co-authorship network, Amazon co-purchase network, YouTube users network, and LiveJournal users network).

Another weakness of our work is that we used only one citation similarity metric (extended direct citation) and only one text similarity metric (BM25). Future work should ideally evaluate multiple citation and text similarity metrics, because different citation metrics and different text metrics may yield different results.

\section{Conclusion}

In this paper we explored science maps of mostly biomedical topics, analyzing the clustering effectiveness for different topic categories. We hope our work will contribute to a more effective use of science maps. We addressed the following research questions:

\paragraph{Which topic categories have the highest and lowest clustering effectiveness in citation and text similarity networks?} We found that the answer is the same for citation and text similarity networks. Paraphrasing the topic category names, the topic categories with the highest clustering effectiveness are diseases, psychology, anatomy, organisms and the techniques and equipment used for diagnostics and therapy, while the topic categories with the lowest clustering effectiveness are natural science fields, geographical entities, information sciences and health care and occupations. Also, the diseases category has a substantially higher clustering effectiveness than all other categories, while the geographical entities category has a substantially lower clustering effectiveness.

\paragraph{Which topic categories have higher clustering effectiveness in citation similarity networks than in text similarity networks, and vice versa?} We found that there are two factors that can make any topic category have higher clustering effectiveness in either network. The first factor is the size of the clusters generated by the clustering process (i.e., the Resolution parameter). The smaller the size, the higher the clustering effectiveness in citation networks relative to text networks. The second factor, specific to our experimental setting, is the percentage of all topic documents that must be covered by the selected clusters (i.e., the Coverage parameter). The higher this percentage, the higher the clustering effectiveness in citation networks relative to text networks. Regardless of these two factors, we found that the diseases and organisms topic categories tend to have higher clustering effectiveness in citation networks than in text networks.

\paragraph{}Our work has shown that there is a strong tendency for clusters in science maps to represent some topics better than others. Further research could explore how to control which topics are clustered better, so that users of science maps can adjust the maps to their needs.

\section{Data availability}

The code used to run the experiments and the data needed to replicate the results are available in Zenodo \cite{bascur_2024_11181030}.

\bibliographystyle{aomplain}
\bibliography{mybib}

\providecommand{\bysame}{\leavevmode\hbox to3em{\hrulefill}\thinspace}
\providecommand{\noopsort}[1]{}
\providecommand{\mr}[1]{\href{http://www.ams.org/mathscinet-getitem?mr=#1}{MR~#1}}
\providecommand{\zbl}[1]{\href{http://www.zentralblatt-math.org/zmath/en/search/?q=an:#1}{Zbl~#1}}
\providecommand{\jfm}[1]{\href{http://www.emis.de/cgi-bin/JFM-item?#1}{JFM~#1}}
\providecommand{\arxiv}[1]{\href{http://www.arxiv.org/abs/#1}{arXiv~#1}}
\providecommand{\doi}[1]{\url{https://doi.org/#1}}
\providecommand{\MR}{\relax\ifhmode\unskip\space\fi MR }
\providecommand{\MRhref}[2]{%
  \href{http://www.ams.org/mathscinet-getitem?mr=#1}{#2}
}
\providecommand{\href}[2]{#2}
\begin{thebibliography}{10}

\bibitem{ahlgren2020enhancing}
\bgroup\scshape{}P.~Ahlgren\egroup{}, \bgroup\scshape{}Y.~Chen\egroup{}, \bgroup\scshape{}C.~Colliander\egroup{}, and \bgroup\scshape{}N.~J. van Eck\egroup{}, Enhancing direct citations: A comparison of relatedness measures for community detection in a large set of pubmed publications,  \emph{Quantitative Science Studies} \textbf{1} no.~2 (2020), 714--729. \doi{10.1162/qss_a_00027}.

\bibitem{bascur_2024_11181030}
\bgroup\scshape{}J.~P. Bascur\egroup{}, {Which topics are best represented by science maps? An analysis of clustering effectiveness for citation and text similarity networks (data)}, June 2024. \doi{10.5281/zenodo.11181030}.

\bibitem{bascur2023academic}
\bgroup\scshape{}J.~P. Bascur\egroup{}, \bgroup\scshape{}S.~Verberne\egroup{}, \bgroup\scshape{}N.~J. van Eck\egroup{}, and \bgroup\scshape{}L.~Waltman\egroup{}, Academic information retrieval using citation clusters: in-depth evaluation based on systematic reviews,  \emph{Scientometrics} \textbf{128} no.~5 (2023), 2895--2921.

\bibitem{chen2017science}
\bgroup\scshape{}C.~Chen\egroup{}, Science mapping: a systematic review of the literature,  \emph{Journal of Data and Information Science} \textbf{2} no.~2 (2017), 1--40. \doi{10.1515/jdis-2017-0006}.

\bibitem{cobo2011science}
\bgroup\scshape{}M.~J. Cobo\egroup{}, \bgroup\scshape{}A.~G. L{\'o}pez-Herrera\egroup{}, \bgroup\scshape{}E.~Herrera-Viedma\egroup{}, and \bgroup\scshape{}F.~Herrera\egroup{}, Science mapping software tools: Review, analysis, and cooperative study among tools,  \emph{Journal of the American Society for Information Science and Technology} \textbf{62} no.~7 (2011), 1382--1402. \doi{10.1002/asi.21525}.

\bibitem{costas2021heterogeneous}
\bgroup\scshape{}R.~Costas\egroup{}, \bgroup\scshape{}S.~de~Rijcke\egroup{}, and \bgroup\scshape{}N.~Marres\egroup{}, “heterogeneous couplings”: Operationalizing network perspectives to study science-society interactions through social media metrics,  \emph{Journal of the Association for Information Science and Technology} \textbf{72} no.~5 (2021), 595--610. \doi{10.1002/asi.24427}.

\bibitem{Fields}
\bgroup\scshape{}CWTS\egroup{}, Leiden ranking fields, 2023, [Online; accessed 20-March-2023]. Available at \url{https://www.leidenranking.com/information/fields}.

\bibitem{ding2011community}
\bgroup\scshape{}Y.~Ding\egroup{}, Community detection: Topological vs. topical,  \emph{Journal of Informetrics} \textbf{5} no.~4 (2011), 498--514. \doi{10.1016/j.joi.2011.02.006}.

\bibitem{van2011methodological}
\bgroup\scshape{}N.~J. van Eck\egroup{}, \emph{Methodological advances in bibliometric mapping of science}, no. EPS-2011-247-LIS, 2011.

\bibitem{fortunato2010community}
\bgroup\scshape{}S.~Fortunato\egroup{}, Community detection in graphs,  \emph{Physics Reports} \textbf{486} no.~3-5 (2010), 75--174. \doi{10.1016/j.physrep.2009.11.002}.

\bibitem{glaser2020opening}
\bgroup\scshape{}J.~Gl{\"a}ser\egroup{}, Opening the black box of expert validation of bibliometric maps,  in \emph{Lockdown Bibliometrics: Papers not submitted to the STI Conference 2020 in Aarhus}, 2020, pp.~27--36.

\bibitem{haunschild2018algorithmically}
\bgroup\scshape{}R.~Haunschild\egroup{}, \bgroup\scshape{}H.~Schier\egroup{}, \bgroup\scshape{}W.~Marx\egroup{}, and \bgroup\scshape{}L.~Bornmann\egroup{}, Algorithmically generated subject categories based on citation relations: An empirical micro study using papers on overall water splitting,  \emph{Journal of Informetrics} \textbf{12} no.~2 (2018), 436--447. \doi{10.1016/j.joi.2018.03.004}.

\bibitem{havemann2017memetic}
\bgroup\scshape{}F.~Havemann\egroup{}, \bgroup\scshape{}J.~Gl{\"a}ser\egroup{}, and \bgroup\scshape{}M.~Heinz\egroup{}, Memetic search for overlapping topics based on a local evaluation of link communities,  \emph{Scientometrics} \textbf{111} (2017), 1089--1118. \doi{10.1007/s11192-017-2302-5}.

\bibitem{held2022know}
\bgroup\scshape{}M.~Held\egroup{}, {Know thy tools! Limits of popular algorithms used for topic reconstruction},  \emph{Quantitative Science Studies} \textbf{3} no.~4 (2022), 1054--1078. \doi{10.1162/qss_a_00217}.

\bibitem{held2020topic}
\bgroup\scshape{}M.~Held\egroup{}, \bgroup\scshape{}G.~Laudel\egroup{}, and \bgroup\scshape{}J.~Gl{\"a}ser\egroup{}, Topic reconstruction from networks of papers may not be possible if only one algorithm is applied to only one data model,  in \emph{Lockdown Bibliometrics: Papers not submitted to the STI Conference 2020 in Aarhus}, 2020, p.~18.

\bibitem{held2021challenges}
\bgroup\scshape{}M.~Held\egroup{}, \bgroup\scshape{}G.~Laudel\egroup{}, and \bgroup\scshape{}J.~Gl{\"a}ser\egroup{}, Challenges to the validity of topic reconstruction,  \emph{Scientometrics} \textbf{126} (2021), 4511--4536. \doi{10.1007/s11192-021-03920-3}.

\bibitem{held2022interpret}
\bgroup\scshape{}M.~Held\egroup{} and \bgroup\scshape{}T.~Velden\egroup{}, {How to interpret algorithmically constructed topical structures of scientific fields? A case study of citation-based mappings of the research specialty of invasion biology},  \emph{Quantitative Science Studies} \textbf{3} no.~3 (2022), 651--671. \doi{10.1162/qss_a_00194}.

\bibitem{hric2014community}
\bgroup\scshape{}D.~Hric\egroup{}, \bgroup\scshape{}R.~K. Darst\egroup{}, and \bgroup\scshape{}S.~Fortunato\egroup{}, Community detection in networks: Structural communities versus ground truth,  \emph{Physical Review E} \textbf{90} no.~6 (2014), 062805. \doi{10.1103/PhysRevE.90.062805}.

\bibitem{klavans2017type}
\bgroup\scshape{}R.~Klavans\egroup{} and \bgroup\scshape{}K.~W. Boyack\egroup{}, Which type of citation analysis generates the most accurate taxonomy of scientific and technical knowledge?,  \emph{Journal of the Association for Information Science and Technology} \textbf{68} no.~4 (2017), 984--998. \doi{10.1002/asi.23734}.

\bibitem{lai2005using}
\bgroup\scshape{}K.-K. Lai\egroup{} and \bgroup\scshape{}S.-J. Wu\egroup{}, Using the patent co-citation approach to establish a new patent classification system,  \emph{Information Processing \& Management} \textbf{41} no.~2 (2005), 313--330. \doi{10.1016/j.ipm.2003.11.004}.

\bibitem{leydesdorff1998theories}
\bgroup\scshape{}L.~Leydesdorff\egroup{}, Theories of citation?,  \emph{Scientometrics} \textbf{43} (1998), 5--25. \doi{10.1007/BF02458391}.

\bibitem{mesh}
\bgroup\scshape{}{National Institutes of Health}\egroup{}, Medical subject headings, Available at \url{https://www.nlm.nih.gov/mesh/meshhome.html}, Accessed: 2023-09-07.

\bibitem{newman2004coauthorship}
\bgroup\scshape{}M.~E. Newman\egroup{}, Coauthorship networks and patterns of scientific collaboration,  \emph{Proceedings of the National Academy of Sciences} \textbf{101} no.~suppl\_1 (2004), 5200--5205. \doi{10.1073/pnas.0307545100}.

\bibitem{newman2016structure}
\bgroup\scshape{}M.~E. Newman\egroup{} and \bgroup\scshape{}A.~Clauset\egroup{}, Structure and inference in annotated networks,  \emph{Nature Communications} \textbf{7} no.~1 (2016), 11863. \doi{10.1038/ncomms11863}.

\bibitem{peel2017ground}
\bgroup\scshape{}L.~Peel\egroup{}, \bgroup\scshape{}D.~B. Larremore\egroup{}, and \bgroup\scshape{}A.~Clauset\egroup{}, The ground truth about metadata and community detection in networks,  \emph{Science Advances} \textbf{3} no.~5 (2017), e1602548. \doi{10.1126/sciadv.1602548}.

\bibitem{INR-019}
\bgroup\scshape{}S.~Robertson\egroup{} and \bgroup\scshape{}H.~Zaragoza\egroup{}, {The Probabilistic Relevance Framework: BM25 and Beyond},  \emph{Foundations and Trends in Information Retrieval} \textbf{3} no.~4 (2009), 333--389. \doi{10.1561/1500000019}.

\bibitem{rossetti2016novel}
\bgroup\scshape{}G.~Rossetti\egroup{}, \bgroup\scshape{}L.~Pappalardo\egroup{}, and \bgroup\scshape{}S.~Rinzivillo\egroup{}, A novel approach to evaluate community detection algorithms on ground truth,  in \emph{Complex Networks VII: Proceedings of the 7th Workshop on Complex Networks CompleNet 2016}, Springer, 2016, pp.~133--144. \doi{10.1007/978-3-319-30569-1_10}.

\bibitem{aggclus}
\bgroup\scshape{}{SAS Institute Inc.}\egroup{}, Clustering methods, Available at \url{https://support.sas.com/documentation/cdl/en/statug/63033/HTML/default/viewer.htm#statug_cluster_sect012.htm}, 2009, Accessed: 2023-09-07.

\bibitem{seitz2021case}
\bgroup\scshape{}C.~Seitz\egroup{}, \bgroup\scshape{}M.~Schmidt\egroup{}, \bgroup\scshape{}N.~Schwichtenberg\egroup{}, and \bgroup\scshape{}T.~Velden\egroup{}, A case study of the epistemic function of citations—implications for citation-based science mapping,  in \emph{Proceedings of the 18th International Conference of the International Society for Scientometrics and Informetrics (ISSI)}, 2021.

\bibitem{sjogaarde2018granularity}
\bgroup\scshape{}P.~Sj{\"o}g{\aa}rde\egroup{} and \bgroup\scshape{}P.~Ahlgren\egroup{}, Granularity of algorithmically constructed publication-level classifications of research publications: Identification of topics,  \emph{Journal of Informetrics} \textbf{12} no.~1 (2018), 133--152. \doi{10.1016/j.joi.2017.12.006}.

\bibitem{sjogaarde2021algorithmic}
\bgroup\scshape{}P.~Sj{\"o}g{\aa}rde\egroup{}, \bgroup\scshape{}P.~Ahlgren\egroup{}, and \bgroup\scshape{}L.~Waltman\egroup{}, Algorithmic labeling in hierarchical classifications of publications: Evaluation of bibliographic fields and term weighting approaches,  \emph{Journal of the Association for Information Science and Technology} \textbf{72} no.~7 (2021), 853--869. \doi{10.1002/asi.24452}.

\bibitem{vsubelj2016clustering}
\bgroup\scshape{}L.~{\v{S}}ubelj\egroup{}, \bgroup\scshape{}N.~J. Van~Eck\egroup{}, and \bgroup\scshape{}L.~Waltman\egroup{}, Clustering scientific publications based on citation relations: A systematic comparison of different methods,  \emph{PLOS One} \textbf{11} no.~4 (2016), e0154404. \doi{10.1371/journal.pone.0154404}.

\bibitem{sullivan2007using}
\bgroup\scshape{}R.~Sullivan\egroup{}, \bgroup\scshape{}S.~Eckhouse\egroup{}, and \bgroup\scshape{}G.~Lewison\egroup{}, Using bibliometrics to inform cancer research policy and spending,  \emph{Monitoring financial flows for health research} (2007), 67--78.

\bibitem{traag2019louvain}
\bgroup\scshape{}V.~A. Traag\egroup{}, \bgroup\scshape{}L.~Waltman\egroup{}, and \bgroup\scshape{}N.~J. Van~Eck\egroup{}, {From Louvain to Leiden: Guaranteeing well-connected communities},  \emph{Scientific Reports} \textbf{9} no.~1 (2019), 5233. \doi{10.1038/s41598-019-41695-z}.

\bibitem{velden2017comparison}
\bgroup\scshape{}T.~Velden\egroup{}, \bgroup\scshape{}K.~W. Boyack\egroup{}, \bgroup\scshape{}J.~Gl{\"a}ser\egroup{}, \bgroup\scshape{}R.~Koopman\egroup{}, \bgroup\scshape{}A.~Scharnhorst\egroup{}, and \bgroup\scshape{}S.~Wang\egroup{}, Comparison of topic extraction approaches and their results,  \emph{Scientometrics} \textbf{111} no.~2 (2017), 1169--1221. \doi{10.1007/s11192-017-2306-1}.

\bibitem{waltman2020principled}
\bgroup\scshape{}L.~Waltman\egroup{}, \bgroup\scshape{}K.~W. Boyack\egroup{}, \bgroup\scshape{}G.~Colavizza\egroup{}, and \bgroup\scshape{}N.~J. van Eck\egroup{}, A principled methodology for comparing relatedness measures for clustering publications,  \emph{Quantitative Science Studies} \textbf{1} no.~2 (2020), 691--713. \doi{10.1162/qss_a_00035}.

\bibitem{waltman2012new}
\bgroup\scshape{}L.~Waltman\egroup{} and \bgroup\scshape{}N.~J. Van~Eck\egroup{}, A new methodology for constructing a publication-level classification system of science,  \emph{Journal of the American Society for Information Science and Technology} \textbf{63} no.~12 (2012), 2378--2392. \doi{10.1002/asi.22748}.

\bibitem{xu2018overlapping}
\bgroup\scshape{}S.~Xu\egroup{}, \bgroup\scshape{}J.~Liu\egroup{}, \bgroup\scshape{}D.~Zhai\egroup{}, \bgroup\scshape{}X.~An\egroup{}, \bgroup\scshape{}Z.~Wang\egroup{}, and \bgroup\scshape{}H.~Pang\egroup{}, Overlapping thematic structures extraction with mixed-membership stochastic blockmodel,  \emph{Scientometrics} \textbf{117} no.~1 (2018), 61--84. \doi{10.1007/s11192-018-2841-4}.

\bibitem{yuan2022measurement}
\bgroup\scshape{}M.~Yuan\egroup{}, \bgroup\scshape{}J.~Zobel\egroup{}, and \bgroup\scshape{}P.~Lin\egroup{}, Measurement of clustering effectiveness for document collections,  \emph{Information Retrieval Journal} \textbf{25} no.~3 (2022), 239--268. \doi{10.1007/s10791-021-09401-8}.

\bibitem{zitt2015meso}
\bgroup\scshape{}M.~Zitt\egroup{}, {Meso-level retrieval: IR-bibliometrics interplay and hybrid citation-words methods in scientific fields delineation},  \emph{Scientometrics} \textbf{102} no.~3 (2015), 2223--2245. \doi{10.1007/s11192-014-1482-5}.

\end{thebibliography}

\end{document}